\documentclass[preprint,aps,floatfix,nofootinbib,preprintnumbers,amsmath, amssymb,showkeys]{revtex4-1}

\usepackage{epsf,epsfig,subfigure,graphicx,amsmath,amssymb}
\usepackage{color}
\usepackage{slashed}
\usepackage{amsmath}
\usepackage{mathtools}
\usepackage{color}

\allowdisplaybreaks

\def\lsim{\mathrel{\rlap{\lower4pt\hbox{\hskip1pt$\sim$}}
		\raise1pt\hbox{$<$}}}         
\def\gsim{\mathrel{\rlap{\lower4pt\hbox{\hskip1pt$\sim$}}
		\raise1pt\hbox{$>$}}}         

\newcommand*\rd[1]{%
	\textcolor{red}{#1}}

\begin{document}
	
	\title{\bf 	Curvature Spinors in Locally Inertial Frame \\
 and the Relations with Sedenion }

\author{I.~K.~ Hong$^1$\footnote{Email at: hijko3@yonsei.ac.kr}\,
C.~S.~ Kim$^{1,2}$\footnote{Email at: cskim@yonsei.ac.kr} and
G.~H.~ Min$^{1}$\footnote{Email at: mk9538@yonsei.ac.kr}\\[5mm]$^1$Department of Physics and IPAP, Yonsei University,  
	Seoul 03722, Korea\\[5mm]$^2$Institute of High Energy Physics, Dongshin University,  
	Naju 58245, Korea}%

	\vspace{2.0cm}
	\begin{abstract}
		In the 2-spinor formalism, the gravity can be dealt with curvature spinors with four spinor indices. Here we show a new effective method to express the components of curvature spinors in the rank-2 $4 \times 4$ tensor representation for the gravity in a locally inertial frame.
		In the process we have developed a few manipulating techniques, through which the roles of each component of Riemann curvature tensor are revealed. We define a new algebra `sedon', whose structure is almost the same as sedenion except the basis multiplication rule.  Finally we also show that curvature spinors can be represented in the sedon form and observe the chiral structure in curvature spinors. A few applications of the sedon representation, which includes the quaternion form of differential Binanchi indentity, are also presented.
\\
	\end{abstract}

	
	\maketitle

	\section{Introduction}
	In the 2-spinor formalism \cite{penrose1984spinors,bain2000coordinate,carmeli2000theory} all tensors with spacetime indices can be transformed into  spinors with twice the number of spinor indices,
i.e.  a rank-2 tensor is changed into a spinor with four spinor indices. In addition,  if the tensor is antisymmetric and real, it can be represented by a sum of two spinors with  two spinor indices, and they are complex conjugate of each other, which indicates that a rank-2 antisymmetric tensor is equivalent to a spinor with two spinor indices.
	The Riemann curvature tensor is a rank-4 real tensor which describes gravitational fields and it has two antisymmetric characters. It means that the gravity can be described by two spinors with four spinor indices.
	Those two spinors are called as curvature spinors: one of them is Ricci spinor and the other is Weyl conformal spinor \cite{penrose1984spinors, penrose1960spinor,o2003introduction,carmeli2000theory}.

At any points on a pseudo-Riemannian manifold, we can find a locally flat coordinate \cite{carroll2004spacetime}, whose metric is Minkowski. Though the metric is locally Minkowski, the second derivative of the metric is not necessarily zero and the Riemann curvature tensor as well as curvature spinors do not have to be zero. Here we can obtain the explicit representations of curvature spinors, whose components  can be easily identified by using new techniques, i.e. manipulating spinor indices and rotating sigma basis in locally flat coordinates \cite{hong2019quaternion}. Then all the components of curvature spinors are represented with simple combinations of Riemann curvature tensors. The obtained representation can be used not only in a speciallt flat coordinate but also for vielbein indices or in any other normal coordinates, like Riemann normal coordinate and Fermi coordinate \cite{klein2008general,marzlin1994physical,chicone2006explicit,nesterov1999riemann,muller1999closed,hatzinikitas2000note,yepez2011einstein,nuastase2019classical,ortin2004gravity}.
	By comparing the final forms of Ricci spinors with the spinor forms of Einstein equation, we are able to figure out the roles of each component of Riemann curvature tensor, whose components serve as momentum, energy or stress of gravitational fields.
	Furthermore, we find that the Weyl sipnor can be analyzed by dividing into real part and pure imaginary part,  henceforth,
the components of Weyl conformal spinor can be represented as a simple combination of Wely tensors in flat coordinate.
	
	There are already a quite few papers that show the relation between sedenion and gravitational field, however, all are restricted to a weak gravitational field in a flat frame \cite{mironov2014sedeonic, kansu2014representation,chanyal2014sedenion,koplinger2007gravity}. Here we express the basis of sedenion as a set of direct product of quaternion basis, through which we can define a new algebra `sedon', whose structure is similar to sedenion  except the basis multiplication rule. We will show that the curvature spinors for general gravitational fields in a locally flat coordinates can be regarded as a sedon.
From the sedon form of curvature spinors, we can get a view of the curvature spinors as the combination of  right-handed and left-handed rotational effects. And we also introduce a few  applications of the sedon form with multiplication techniques. One of the application is the quaternion form of differential Bianchi identity and, in the process, we introduce a new index notation with the spatially opposite-handed quantities.

	\section{Tensor Representation of a Field with two spinor Indices }

 In this section we introduce the basics about the 2-spinor formalism,  which are explained  in detail in our earlier paper \cite{hong2019quaternion}.

 Any tensor   $T_{abc..} $ with four dimensional spacetime indices $a,b,c,..$, can be inverted into a spinor with spinor indices $ A, A',B, B',..$ like $T_{AA'BB'..}$ by multiplying Infeld-van der Waerden symbols $g^{\quad a}_{AA'}$, 
	\begin{eqnarray}
	T_{AA'BB'..}=T_{ab..} g^{\quad a}_{AA'} g^{\quad a}_{BB'} ..  \quad .
	\end{eqnarray}
	In Minkowski spacetime, $ g^{\quad a}_{AA'} $ is $ \frac{1}{\sqrt{2}} \sigma^{a}_{\;\; AA'}$, where $\sigma^{a}_{\;\; AA'}$ are  four-sigma matrices  $(\sigma^{0},\sigma^{1},\sigma^{2},\sigma^{3})$; $\sigma^0$ is $2 \times 2$ identity matrix and $\sigma^1, \sigma^2, \sigma^3$ are Pauli matrices. Eq. (1)  can be written conventionally as
	\begin{eqnarray}
	T_{AA'BB'..}=T_{ab..}    .
	\end{eqnarray}

	Any arbitrary anti-symmetric tensor $F_{ab}=F_{AA'BB'}$ can be expressed as the sum of two symmetric spinors as
	\begin{eqnarray}
	F_{AA'BB'}=\varphi_{AB} \varepsilon_{A'B'} +\varepsilon_{AB} \psi_{A'B'},   \label{F}
	\end{eqnarray}
	where $\varphi_{AB} =\frac{1}{2} F_{ABC'}^{\qquad C'}$ and $\psi_{A'B'}=\frac{1}{2} F_{C\;\;A'B'}^{\;\; C} $ are symmetric spinors	(unprimed and primed spinor indices can be switched back and forth each other), and $\varepsilon^{AB}$, $\varepsilon^{A'B'}$, $\varepsilon_{AB}$, $\varepsilon_{A'B'}$ are the $\varepsilon$-spinors whose components are $\varepsilon^{12}=\varepsilon_{12}=+1, \varepsilon^{21}=\varepsilon_{21}=-1 $  \cite{penrose1984spinors}; $k^A=\varepsilon^{AB}k_B,k_B=k^A\varepsilon_{AB}$.
	If $F_{ab}$ is real, then $\psi_{A'B'}=\bar{\varphi}_{A'B'}$ (where $\bar{\varphi}$ is the
complex conjugate of ${\varphi}$) and
	\begin{eqnarray}F_{ab}=F_{AA'BB'}=\varphi_{AB} \varepsilon_{A'B'} +\varepsilon_{AB} \bar{\varphi}_{A'B'}  . \label{at}
	\end{eqnarray}
		We have shown the components of $\varphi_{AB}$ and $\bar{\varphi}_{A'B'}$ explicitly in flat spacetime in \cite{hong2019quaternion}. The sign conventions for the Minkowski metric is $g_{\mu\nu} = \rm{diag}(1,-1,-1,-1)$.

For any real anti-symmetric tensor $F_{AA'BB'}$, we can write  as
	\begin{eqnarray}
	F_{AA'BB'}=\frac{1}{2} F_{\mu\nu} \sigma^{\mu}_{AA'} \sigma^{\nu}_{BB'} =\frac{1}{2} F_{\mu\nu} \sigma^{\mu}_{AA'} \bar{\sigma}^{\nu\; C'C} \varepsilon_{C'B'} \varepsilon_{CB},
	\end{eqnarray}
	where $\bar{\sigma}^{\mu}=(\sigma^{0},-\sigma^{1},-\sigma^{2},-\sigma^{3})$,
	then
	\begin{eqnarray}
	\varphi_{AB}&&=\frac{1}{2}F_{AA'B}^{\qquad A'}
	=\frac{1}{2}F_{AA'BB'}\varepsilon^{A'B'}
	\nonumber \\&&=\frac{1}{4}F_{\mu\nu} \sigma^{\mu}_{AA'} \bar{\sigma}^{\nu \; C'C} \varepsilon_{C'B'} \varepsilon_{CB}\varepsilon^{A'B'}
	=\frac{1}{4}F_{\mu\nu} \; \sigma^{\mu}_{AA'} \bar{\sigma}^{\nu\; A'C}  \varepsilon_{CB}~,
	\\
	 \bar{\varphi}_{A'B'}   &&=\frac{1}{2}F_{AA'\;B'}^{\quad A}
	=\frac{1}{2}F_{AA'BB'}\varepsilon^{AB} \nonumber
	\\&&=\frac{1}{4}F_{\mu\nu} \bar{\sigma}^{\mu C'C} \sigma^{\nu}_{BB'} \varepsilon_{A'C'} \varepsilon_{AC}\varepsilon^{AB}
	=\frac{1}{4}F_{\mu\nu} \;\varepsilon_{A'C'} \bar{\sigma}^{\mu C'B} \sigma^{\nu}_{BB'}~.
	\end{eqnarray}
Since
	\begin{eqnarray}
	\sigma^{\mu}_{AA'} \bar{\sigma}^{\nu \; A'C}=
	\begin{pmatrix}
	\sigma^{0}\sigma^{0} & -\sigma^{0}\sigma^{1} & -\sigma^{0}\sigma^{2} & -\sigma^{0}\sigma^{3} \\
	\sigma^{1}\sigma^{0} & -\sigma^{1}\sigma^{1} & -\sigma^{1}\sigma^{2} & -\sigma^{1}\sigma^{3} \\
	\sigma^{2}\sigma^{0} & -\sigma^{2}\sigma^{1} & -\sigma^{2}\sigma^{2} & -\sigma^{2}\sigma^{3} \\
	\sigma^{3}\sigma^{0} & -\sigma^{3}\sigma^{1} & -\sigma^{3}\sigma^{2} & -\sigma^{3}\sigma^{3}
	\end{pmatrix}_{A  {\raisebox{ 8  pt}{$\scriptstyle C$}}  }=
	\begin{pmatrix}
	\sigma^{0} & -\sigma^{1} & -\sigma^{2} & -\sigma^{3} \\
	\sigma^{1} & -\sigma^{0} & -i\sigma^{3}  & i\sigma^{2}  \\
	\sigma^{2} & i\sigma^{3} &- \sigma^{0} & -i\sigma^{1} \\
	\sigma^{3} & - i\sigma^{2} & i\sigma^{1} & -\sigma^{0}
	\end{pmatrix}_{A  {\raisebox{ 8  pt}{$\scriptstyle C$}}}, \label{80}
	\end{eqnarray}
	$\varphi_{A}^{\;\;\; D}= \varepsilon^{DB}\varphi_{AB} $ becomes
	\begin{eqnarray}
	\varphi_{A}^{\;\;\; D}\;&&=\frac{1}{4} F_{\mu\nu} \sigma^{\mu}_{AA'} \bar{\sigma}^{\nu \; A'D} \nonumber \\&&= \frac{1}{4} \left [ \begin{pmatrix}
	0 & -F_{10} & -F_{20} & -F_{30} \\
	F_{10} & 0 & F_{12} & F_{13} \\
	F_{20} & -F_{12} & 0 & F_{23} \\
	F_{30} & -F_{13} & -F_{23} & 0
	\end{pmatrix}
	\begin{pmatrix}
	\sigma^{0} & -\sigma^{1} & -\sigma^{2} & -\sigma^{3} \\
	\sigma^{1} & -\sigma^{0} & -i\sigma^{3}  & i\sigma^{2}  \\
	\sigma^{2} & i\sigma^{3} &- \sigma^{0} & -i\sigma^{1} \\
	\sigma^{3} & - i\sigma^{2} & i\sigma^{1} &- \sigma^{0}
	\end{pmatrix} ^T\; \right ]_{A  {\raisebox{  8 pt}{$\scriptstyle D$}}  }\nonumber \\&&
	=\frac{1}{2}(F_{i0}\sigma^{i}-\frac{1}{2}i\,\epsilon_{\;\;k}^{ij}F_{ij}\sigma^k )_A^{\;\;D},
	\label{86}
	\end{eqnarray}
	where $i\;,j\;,k$ are the 3-dimensional vector indices which have the value 1, 2 or 3, and $\epsilon_{\;\;k}^{ij}$ is $\epsilon_{pqk} \delta_p^i \delta_q^j$ for the Levi-Civita symbol $\epsilon_{ijk}$. Einstein summation convention is used for 3-dimensional vector indices $i,j$ and $k$.
	Similar to (\ref{80}) and (\ref{86}),
	\begin{eqnarray}
	\bar{\sigma}^{\mu C'B} \sigma^{\nu}_{BB'}=
	\begin{pmatrix}
	\sigma^{0}\sigma^{0} & \sigma^{0}\sigma^{1} & \sigma^{0}\sigma^{2} & \sigma^{0}\sigma^{3} \\
	-\sigma^{1}\sigma^{0} & -\sigma^{1}\sigma^{1} & -\sigma^{1}\sigma^{2} & -\sigma^{1}\sigma^{3} \\
	-\sigma^{2}\sigma^{0} & -\sigma^{2}\sigma^{1} & -\sigma^{2}\sigma^{2} & -\sigma^{2}\sigma^{3} \\
	-\sigma^{3}\sigma^{0} & -\sigma^{3}\sigma^{1} & -\sigma^{3}\sigma^{2} & -\sigma^{3}\sigma^{3}
	\end{pmatrix} _{{\raisebox{  8 pt}{$\scriptstyle C'$}B'}}\!\!=\begin{pmatrix}
	\sigma^{0} & \sigma^{1} & \sigma^{2} & \sigma^{3} \\
	-\sigma^{1} & -\sigma^{0} & -i\sigma^{3}  & i\sigma^{2}  \\
	-\sigma^{2} & i\sigma^{3} & -\sigma^{0} & -i\sigma^{1} \\
	-\sigma^{3} & - i\sigma^{2} & i\sigma^{1} & -\sigma^{0}
	\end{pmatrix}_{{\raisebox{  8 pt}{$\scriptstyle C'$}B'}} \!\!\!,
	\end{eqnarray}
	\begin{eqnarray}
	\bar{\varphi}^{D'}_{\;\;B'}=\varepsilon^{D'A'}\bar{\varphi}_{A'B'}=-\frac{1}{4}F_{\mu\nu} \;\bar{\sigma}^{\mu D'B} \sigma^{\nu}_{BB'}
	= \frac{1}{2}(F_{i0}\sigma^{i}+\frac{1}{2}i\, \epsilon_{\;\;k}^{ij} F_{ij}\sigma^k )^{D'}_{\;\; B'}. \label{88}
	\end{eqnarray}
	If we denote matrix representation of $\varepsilon_{AB}$ by $\mathbb{\varepsilon}$, then
	\begin{eqnarray}
	\mathbb{\sigma}^\mu\mathbb{\varepsilon}=(\sigma^{0},\sigma^{1},\sigma^{2},\sigma^{3})\mathbb{\varepsilon}=(i\sigma^{2},-\sigma^{3},i\sigma^{0},\sigma^{1}),\\
	\mathbb{\varepsilon}\mathbb{\sigma}^\mu=\mathbb{\varepsilon}(\sigma^{0},\sigma^{1},\sigma^{2},\sigma^{3})=(i\sigma^{2},\sigma^{3},i\sigma^{0},-\sigma^{1}).
	\end{eqnarray}
	Let us define $s^\mu$ and $\bar{s}^\mu$ as
	\begin{eqnarray}
	&&s^0=i\sigma^{2}, \quad s^1=-\sigma^{3}, \quad s^2=i\sigma^{0}, \quad s^3=\sigma^{1},  \label{ds} \\
		&&\bar{s}^0=i\sigma^{2}, \quad \bar{s}^1=-\sigma^{3}, \quad \bar{s}^2=-i\sigma^{0}, \quad \bar{s}^3=\sigma^{1},  \label{dsb}
	\end{eqnarray}
	where $\bar{s}^\mu$ is complex conjugate of $s^\mu$.
	Then
	\begin{eqnarray}
	&&\mathbb{\sigma}^\mu\mathbb{\varepsilon}=(s^0, s^1,s^2,s^3),
	\\ &&\mathbb{\varepsilon}\mathbb{\sigma}^\mu=(s^0, -s^1,s^2,-s^3)=(\bar{s}^0, -\bar{s}^1,-\bar{s}^2,-\bar{s}^3),
		\end{eqnarray}
	and
	\begin{eqnarray}
	\varphi_{AB}&&=\varphi_{A}^{\;\;\; D} \varepsilon_{DB}=
	\frac{1}{2}(F_{i0} s^{i}-\frac{1}{2}i\,\epsilon_{\;\;k}^{ij}F_{ij} s^k), 	\label{p1} \\
	\varphi_{A'B'}&&=\varphi^{D'}_{\;\;\; B'} \varepsilon_{D'A'}= -\varepsilon_{A'D'} \phi^{D'}_{\;\;\; B'}
	=\frac{1}{2}(F_{i0} \bar{s}^{i}+\frac{1}{2}i\,\epsilon_{\;\;k}^{ij}F_{ij} \bar{s}^k ) ,	\label{p2}
	\end{eqnarray}
	where $s^i$ have unprimed indices $s^i=s^i_{AB}$ and $\bar{s}^i$ have primed indices $\bar{s}^i=\bar{s}^i_{A'B'}$.

	\section{Einstein field equations and curvature spinors}

		In this section we introduce the basics about general relativity in the 2-spinor formalism,
and flat coordinates on the pseudo-Riemannian manifold.
	For a (torsion-free) Riemann curvature tensor
	 \begin{eqnarray}
	 R^{\mu}_{ \;\; \nu \rho \sigma} = \partial_\rho \Gamma^\mu_{\;\; \nu \sigma}-\partial_\sigma  \Gamma^\mu_{\;\; \nu \rho}
	+\Gamma^\lambda_{\;\; \nu \sigma} \Gamma^\mu_{\;\; \lambda \rho}-\Gamma^\lambda_{\;\; \nu \rho} \Gamma^\mu_{\;\; \lambda \sigma},
	\end{eqnarray}
	where $ \Gamma^\rho_{\;\; \mu \nu}$	is a Chistoffel symbol
	\begin{eqnarray}
	\Gamma^\rho_{\;\; \mu \nu}= \frac{1}{2} g^{\rho \lambda} (\partial_\mu g_{ \nu \lambda}+\partial_\nu g_{ \mu \lambda}-\partial_\lambda g_{ \mu \nu}).
	\end{eqnarray}
	Here $R_{\mu \nu \rho \sigma}$ has follwing properties \cite{carroll2004spacetime}
	\begin{eqnarray}
	R_{\mu \nu \rho \sigma}=-R_{ \nu \mu \rho \sigma}~, \label{22}\\
	R_{\mu \nu \rho \sigma}=-R_{\mu \nu \sigma \rho }~, \label{23}\\
	R_{\mu \nu \rho \sigma}=R_{ \rho \sigma \mu \nu}~.
	\end{eqnarray}
	In short, we can denote as
	\begin{eqnarray}
	R_{\mu \nu \rho \sigma}=R_{([\mu \nu] [\rho \sigma])},  \label{prop}  \end{eqnarray}
	where parentheses (~) and square brackets [~] indicates symmetrization and anti-symmetrization of the indices \cite{carroll2004spacetime}. The Riemann curvature tensor  has two kinds of Bianchi identities
	 	 \begin{eqnarray}
	 	 R_{\mu [\nu \rho \sigma]}=0, \label{B1}\\
	 	 \nabla_{[\lambda}R_{\mu \nu] \rho \sigma}=0\label{B2},
	 	 \end{eqnarray}
		where  $\nabla_{\lambda} A^\mu=\partial_\lambda A^\mu + \Gamma^\mu _{\nu \lambda} A^\nu$.

		From the antisymmetric properties of Riemann curvature tensor, it can be decomposed into sum of curvature spinors, $X_{ABCD}$ and $\Phi_{ABC'D'}$, as
	\begin{eqnarray}
	R_{abcd}&&=\frac{1}{2} R_{AX'B\quad cd}^{\qquad X'} \varepsilon_{A'B'}+ \frac{1}{2} R_{XA'\;\;B' \;\;cd}^{\quad \; X} \varepsilon_{AB} \nonumber
	\\&&=\Phi_{ABC'D'} \epsilon_{A'B'}\epsilon_{CD}+\bar{\Phi}_{A'B'CD} \epsilon_{AB}\epsilon_{C'D'} +X_{ABCD}\epsilon_{A'B'}\epsilon_{C'D'} +\bar{X}_{A'B'C'D'} \epsilon_{AB}\epsilon_{CD},
	\label{R}
	\end{eqnarray}
	where
	\begin{eqnarray}
	X_{ABCD}=R_{AX'B\quad CY'D}^{\qquad X'\qquad Y'}, \qquad
	\Phi_{ABC'D'}=R_{AX'B\quad YC'\;\;D'}^{\qquad X'\quad \;\;Y}.
	\end{eqnarray}
The totally symmetric part of $X_{ABCD}$
 \begin{eqnarray}
 \Psi_{ABCD}=X_{A(BCD)}=X_{(ABCD)}
 \end{eqnarray}	
 is called gravitational spinor or Weyl conformal spinor, and $\Phi_{ABC'D'}$ is referred as Ricci spinor
 \cite{penrose1984spinors,penrose1960spinor, o2003introduction}.
	It is well known that
	\begin{eqnarray}
	\Phi_{AA'BB'}=\Phi_{ab}=\Phi_{ba}=\bar{\Phi}_{ab}, \qquad \Phi^{\;\;a}_{a}=0,
	\end{eqnarray}
	and Einstein tensor is
	\begin{eqnarray}
	G_{ab}= R_{ab}-\frac{1}{2}R g_{ab}=-\Lambda g_{ab}-2\Phi_{ab} \label{29},
	\end{eqnarray}
	where
	$\Lambda=X_{AB}^{\quad AB}$, which is equal to $R/4$ \cite{penrose1984spinors}.
	Therefore, the Einstein field equation
	\begin{eqnarray}
	G_{ab}+\lambda g_{ab}=8\pi G T_{ab},  \label{30}
	\end{eqnarray}
	 where $\lambda$ is a cosmology constant, can be written in the form
	\begin{eqnarray}
	\Phi_{ab}=4\pi G(-T_{ab}+\frac{1}{4} T^q_q g_{ab}), \qquad \Lambda=-2 \pi GT^q_q+\lambda . \label{EQ}
	\end{eqnarray}
	Since any symmetric tensor $U_{ab}$ can be expressed as
	\begin{eqnarray}
	U_{ab}=U_{AA'BB'}=S_{ABA'B'}+\varepsilon_{AB}\varepsilon_{A'B'} \tau,
	\end{eqnarray}
	where $\tau=\frac{1}{4} T_c^c$ and $S_{ABA'B'}$ is traceless and symmetric \cite{penrose1984spinors},
	the traceless part of the energy-momentum (symmetric) tensor $T_{ab}$ can be written by $S_{ab}=T_{ab}-\frac{1}{4}T_c^cg_{ab}$. Therefore,  the spinor form of Einstein equations (\ref{EQ})
becomes
	\begin{eqnarray}
	\Phi_{ABA'B'}=-4\pi G S_{ABA'B'}, \qquad X_{AB}^{\quad AB}=-8\pi G \tau + \lambda.
	\end{eqnarray}

		Weyl tensor $C_{\mu \nu \rho \sigma }$ which is another measure of the curvature of spacetime, like Riemann curvature tensor, is defined as \cite{carroll2004spacetime,wald1984general}
		\begin{eqnarray}
		C_{\mu \nu \rho \sigma }
		=R_{\mu \nu \rho \sigma }+\frac{1}{2}(R_{\mu \sigma }g_{\nu \rho }-R_{\mu \rho }g_{\nu \sigma }+
		R_{\nu \rho }g_{\mu \sigma }-R_{\nu \sigma }g_{\mu \sigma })
		+\frac{1}{6}R(g_{\mu \rho }g_{\sigma \nu }-g_{\mu \sigma }g_{\nu \rho }).  \label{C}
		\end{eqnarray}
		It has the same propterties with (\ref{22}), (\ref{23}) and (\ref{B1}).
	It is known \cite{penrose1984spinors} that Weyl tensor has the following relationship with Weyl conformal spinor $\Psi_{ABCD}$:
	\begin{eqnarray}	C_{abcd}=\Psi_{ABCD}\varepsilon_{A'B'}\varepsilon_{C'D'}+\bar{\Psi}_{A'B'C'D'}\varepsilon_{AB}\varepsilon_{CD}.
	\end{eqnarray} 	
	
	At any point $P$ on the pseudo-Riemannian manifold, we can find a flat coordinate system, such that,
	\begin{eqnarray}
	g_{\mu \nu}(P)=\eta_{\mu \nu}, \qquad \frac{\partial g_{\mu\nu}}{\partial x^\lambda} \bigg\rvert_P=0,
		\end{eqnarray}
		where $g_{\mu \nu}(P)$ is the metric at the point $P$ and $\eta_{\mu \nu}$ is the Minkowski metric.
	In this coordinate system, while the Christoffel symbol is zero, the Riemann curvature tensor is \cite{schutz2009first,foster2010short}
	\begin{eqnarray}
	R_{\mu \nu \rho \sigma}= \frac{1}{2} (\partial_\nu \partial_\rho  g_{ \mu \sigma   } -\partial_\nu \partial_\sigma  g_{ \mu \rho  }+\partial_\mu \partial_\sigma  g_{ \nu  \rho  }-\partial_\mu \partial_\rho  g_{ \nu \sigma  }).
		\end{eqnarray}
For future use we introduce Fermi coordinate, which is one of the locally flat coordinate whose time axis is a tangent of a geodesic.  The coordinate follows the Fermi conditions
		\begin{eqnarray}
		g_{\mu \nu}|_G=\eta_{\mu\nu}, 	\qquad \Gamma^\rho_{\mu\nu}|_G=0,
		\end{eqnarray}
along the geodeic G.

	\section{The Tensor representation of Curvature Spinors}

			In this section we  show the process of representing curvature spinors in 4 $\times$ 4 matrices or 3 $\times$ 3 matrices. And we discuss physical implications of those representations.
From now on, we will always use locally flat coordinate for spacetime indices, and use small letters $i,j...z$ as a three dimensional indices, which can be $1,2$ or $3$; while small letters $a,b...h$ as a four dimensional indices, which can be $0,1,2$ or $3$.
	From (\ref{at}) and (\ref{R}), we can lead to
	\begin{eqnarray}
	R_{abcd}
	&&=\phi_{AB,cd} \varepsilon_{A'B'}+\varepsilon_{AB}\bar{\phi}_{A'B',cd} \nonumber\\ &&=\Phi_{ABC'D'} \epsilon_{A'B'}\epsilon_{CD}+\bar{\Phi}_{A'B'CD} \epsilon_{AB}\epsilon_{C'D'} +X_{ABCD}\epsilon_{A'B'}\epsilon_{C'D'} +\bar{X}_{A'B'C'D'} \epsilon_{AB}\epsilon_{CD} ,
	\end{eqnarray}
	where
	\begin{eqnarray}
	\phi_{AB,cd}=\frac{1}{2}(R_{i0\;cd}s^i-\frac{1}{2}i\,\epsilon_{ijk}R_{ij\;cd}s^k ),\\
	\bar{\phi}_{A'B',cd}=\frac{1}{2}(R_{i0\;cd}\bar{s}^i+\frac{1}{2}i\,\epsilon_{ijk}R_{ij\;cd}\bar{s}^k ),
	\end{eqnarray}
	from (\ref{p1}) and (\ref{p2}). We write here the form of $\epsilon_{\;\;k}^{ij}$ as $\epsilon_{ijk}$ for convenience; it is not so difficult to recover the upper- and lower-indices.
	By decomposing $\phi_{AB,cd}$ one more times, we get
		\begin{eqnarray}
	\Phi_{ABC'D'}
	&&=\frac{1}{4} (R_{i0\;j0}s^i \bar{s}^j+\frac{1}{2} i\epsilon_{pqr}R_{i0\;pq}s^i \bar{s}^r -\frac{1}{2} i \epsilon_{ijk} R_{ij\;l0} s^k \bar{s}^l  + \frac{1}{4} \epsilon_{ijk} \epsilon_{pqr} R_{ij\;pq} s^k \bar{s}^r), \nonumber \\
	&&=\frac{1}{4}(R_{k0\;l0}+\frac{1}{2} i\epsilon_{pql}R_{k0\;pq} -\frac{1}{2} i \epsilon_{ijk} R_{ij\;l0}  + \frac{1}{4} \epsilon_{ijk} \epsilon_{pql} R_{ij\;pq}) s^k \bar{s}^l, \label{P0}
	\end{eqnarray}
	\begin{eqnarray}
	X_{ABCD}
	&&=\frac{1}{4}(R_{0i\;0j}s^is^j-\frac{1}{2} i\epsilon_{pqr}R_{0i\;pq}s^is^r -\frac{1}{2} i \epsilon_{ijk} R_{ij\;0l} s^k s^l  - \frac{1}{4} \epsilon_{ijk} \epsilon_{pqr} R_{ij\;pq} s^k s^r) \nonumber  \\
	&&=\frac{1}{4}(R_{0k\;0l}-\frac{1}{2} i\epsilon_{pql}R_{0k\;pq} -\frac{1}{2} i \epsilon_{ijk} R_{ij\;0l}  - \frac{1}{4} \epsilon_{ijk} \epsilon_{pql} R_{ij\;pq}) s^k s^l. \label{X}
	\end{eqnarray}
 We note that $\Phi$ and $X$ are expressed with two 3-dimensional basis like the form in 3$\times$3 basis. Even though there is no 0-th base, which may be related to the curvature of time, $\Phi$ and $X$ can fully describe the spacetime structure.
If $\Phi$ and $X$ are represented in Fermi coordinate, the disappearance of 0-th compomonents of the $s^i$ basis may come from the fact that time follows proper time. However, since here (\ref{P0}) and (\ref{X}) are expressed not only in Fermi coordinate but also in general locally flat coordinates, the representations (\ref{P0}) and (\ref{X}) may demand a new interpretation of time, which is not just as a component of fourth axis in 4-dimension.

	We can divide (\ref{P0}) into two terms by defining
	\begin{eqnarray}
	&&P_{ij} \equiv \frac{1}{2}  \epsilon_{pqj} R_{i0\;pq} -\frac{1}{2} \epsilon_{pqi}R_{j0\;pq}, \label{40}\\
	&&S_{ij}\equiv R_{0i\;0j} + \frac{1}{4} \epsilon_{pqi} \epsilon_{rsj} R_{pq\;rs} , \label{41}
	\\ && \Theta_{ij}\equiv R_{i0\;j0}+\frac{1}{2} i\epsilon_{pqj}R_{i0\;pq} -\frac{1}{2} i \epsilon_{pqi} R_{pq\;j0}  + \frac{1}{4} \epsilon_{pqi} \epsilon_{rsj} R_{pq\;rs}=S_{ij}+i\,P_{ij} \label{S},
	\end{eqnarray}
	where $P_{ij}$ is anti-symmetric and $S_{ij}$ is symmetric for $i,j$. Then (\ref{P0}) is represented as
	\begin{eqnarray}
	\Phi_{ABC'D'}= \frac{1}{4} \Theta_{ij}s^i_{AB}\bar{s}^j_{C'D'} . \label{phi}
	\end{eqnarray}
		The components of $P_{ij}$ and $S_{ij}$ can be simply expressed as
				\begin{eqnarray}
		-\frac{1}{2} \epsilon_{ijk}P_{ij} &&= -\frac{1}{4} (\epsilon_{ijk} \epsilon_{pqj} R_{i0\;pq} - \epsilon_{ijk} \epsilon_{pqi}R_{j0\;pq})= R_{0i\;ki}~,   \label{ep}\\
		S_{\underline{i}\,\underline{j}}&&=R_{0\underline{i}\;0\underline{j}}+ \varepsilon_{\underline{i}\,\underline{p}\underline{q}}\varepsilon_{\underline{j}\,\underline{r}\underline{s}} R_{\underline{p}\underline{q}\;\underline{r}\underline{s}}~,\\
		S_{\underline{i}\,\underline{i}}&&=R_{0\underline{i}\;0\underline{i}}+ |\varepsilon_{\underline{i}\,\underline{p}\underline{q}}| R_{\underline{p}\underline{q}\;\underline{p}\underline{q}}~,  \label{Sii}
		\end{eqnarray}
		where the underlined symbols in subscripts are the value-fixed indices which does not sum up for dummy indices; one of example is $S_{11}=R_{01\;01}+R_{23\;23}$.

	We can express $\Phi_{ABCD}$ as a tensor by multiplying $ g_\mu^{\;AC'}$, which is
	\begin{eqnarray}
	g_\mu^{\;AB'}=\varepsilon^{AC}\varepsilon^{B'D'} g_{\mu\nu} g^\nu_{\;CD'}=\frac{1}{\sqrt{2}}(\sigma^0,\sigma^1,-\sigma^2,\sigma^3)^{AB'}=\frac{1}{\sqrt{2}}\sigma^{t\;\;AB'}_{\; \mu}= \frac{1}{\sqrt{2}}\sigma^{*\;\;AB'}_{\; \mu}~,
	\end{eqnarray}
	where sigma matrices with superscript $\sigma^t$ and $ \sigma^*$ mean the transpose and the complex conjugate of $\sigma$.
	To calculate $\Phi_{ABC'D'} g_\mu^{\;AC'} g_\nu^{\;BD'}=(1/4\,\Theta_{ij}s^i_{AB}\bar{s}^j_{C'D'}) g_\mu^{\;AC'} g_\nu^{\;BD'}$,
	let us define
	\begin{eqnarray}
	f(k,l)_{\mu \nu}=\sigma^k_{AB} \sigma^l_{C'D'}  g_\mu^{\;AC'} g_\nu^{\;BD'}=\frac{1}{2} (\sigma^k_{AB} \sigma^l_{C'D'} ) \sigma_{\;\mu}^{*\;\;AC'} \sigma_{\;\nu}^{*\;BD'}.
	\end{eqnarray}
	Values of $f(k,l)_{\mu \nu}$ are shown in Table \ref{table1}.
	\begin{table}[]
		\begin{tabular}{ccc}
			\hline
			$f(0,1)_{\mu \nu}=\begin{pmatrix}
			0 & 1 & 0 & 0 \\
			1 & 0 & 0 & 0 \\
			0 & 0 & 0 &  i \\
			0 & 0 & i & 0
			\end{pmatrix},$
			&
			$f(1,0)_{\mu \nu}=
			\begin{pmatrix}
			0 & 1 & 0 & 0 \\
			1 & 0 & 0 & 0 \\
			0 & 0 & 0 &  -i \\
			0 & 0 & -i & 0
			\end{pmatrix},$
			&
			$f(3,1)_{\mu \nu}=
			\begin{pmatrix}
			0 & 0 & i & 0 \\
			0 & 0 & 0 & 1 \\
			i & 0 & 0 & 0 \\
			0 & 1 & 0 & 0
			\end{pmatrix},$
			\\
			$f(1,3)_{\mu \nu}=
			\begin{pmatrix}
			0 & 0 & -i & 0 \\
			0 & 0 & 0 & 1 \\
			-i & 0 & 0 & 0 \\
			0 & 1 & 0 & 0
			\end{pmatrix} ,$
			&
			$f(0,3)_{\mu \nu}=
			\begin{pmatrix}
			0 & 0 & 0 & 1 \\
			0 & 0 & -i & 0 \\
			0 & -i & 0 & 0 \\
			1 & 0 & 0 & 0
			\end{pmatrix} ,$
			&
			$f(3,0)_{\mu \nu}=
			\begin{pmatrix}
			0 & 0 & 0 & 1 \\
			0 & 0 & i & 0 \\
			0 & i & 0 & 0 \\
			1 & 0 & 0 & 0
			\end{pmatrix},$
			\\
			$f(0,0)_{\mu \nu}=
			\begin{pmatrix}
			1 & 0 & 0 & 0 \\
			0 & 1 & 0 & 0 \\
			0 & 0 & -1 & 0 \\
			0 & 0 & 0 & 1
			\end{pmatrix} ,$
			&
			$f(1,1)_{\mu \nu}=
			\begin{pmatrix}
			1 & 0 & 0 & 0 \\
			0 & 1 & 0 & 0 \\
			0 & 0 & 1 & 0 \\
			0 & 0 & 0 & -1
			\end{pmatrix} ,$
			&
			$f(3,3)_{\mu \nu}=
			\begin{pmatrix}
			1 & 0 & 0 & 0 \\
			0 & -1 & 0 & 0 \\
			0 & 0 & 1 & 0 \\
			0 & 0 & 0 & 1
			\end{pmatrix} $ .\\ \hline  
					\end{tabular}
		\caption{The lists of $f(k,l)_{\mu \nu}=\sigma^k_{AB} \sigma^l_{C'D'}  g_\mu^{\;AC'} g_\nu^{\;BD'}$.} \label{table1}
	\end{table}
	Using this table, we get 4 $\times$ 4 representation of $\Phi_{ABC'D'}$ as
	\begin{eqnarray}
	&&\Phi_{\mu \nu}= \Phi_{ABC'D'} g_\mu^{\;AC'} g_\nu^{\;BD'}\nonumber\\
	&&= \frac{1}{4} \begin{bmatrix} (\Theta_{12} s^1 \bar{s}^2+ \Theta_{21} s^2 \bar{s}^1)\qquad \qquad \qquad
	\\\qquad   +(\Theta_{23} s^2 \bar{s}^3+ \Theta_{32} s^3 \bar{s}^2)\qquad \qquad \qquad
	\\\qquad \qquad +(\Theta_{31} s^3 \bar{s}^1+ \Theta_{13} s^1 \bar{s}^3)\qquad \qquad
	\\\qquad \qquad \qquad \qquad
	+(\Theta_{11} s^1 \bar{s}^1+\Theta_{22} s^2 \bar{s}^2+\Theta_{33} s^3 \bar{s}^3)  \end{bmatrix} _{ABC'D'} g_\mu^{\;AC'} g_\nu^{\;BD'}\nonumber\\
	&&= \frac{1}{4} \begin{bmatrix} (i \Theta_{12} f(3,0)-i\Theta_{21} f(0,3))\qquad \qquad \qquad
	\\\qquad+(i \Theta_{23} f(0,1)-i\Theta_{32} f(1,0) ) \qquad \qquad \qquad
	\\\qquad\qquad + ( -\Theta_{31} f(1,3)-\Theta_{13} f(3,1)) \qquad \qquad
	\\\qquad\qquad\qquad\qquad+(\Theta_{11} f(3,3) +\Theta_{22} f(0,0) +\Theta_{33} f(1,1)) \end{bmatrix}_{\mu\nu}
	\nonumber\\
	&&= \frac{1}{2}\begin{pmatrix}
	\frac{1}{2}(S_{11} \!+\!S_{22}\!+\!S_{33}) & -P_{23} & -P_{31} & -P_{12} \\
	-P_{23} & \frac{1}{2}(-S_{11}\!+\!S_{22}\!+\!S_{33}) & -S_{12} & -S_{31} \\
	-P_{31} & -S_{12} & \frac{1}{2}(S_{11}\!-\!S_{22}\!+\!S_{33}) & -S_{32} \\
	-P_{12} & -S_{31} & -S_{32} & \frac{1}{2}(S_{11}\!+\!S_{22}\!-\!S_{33})
	\end{pmatrix}, \label{P}
	\end{eqnarray}
	which is a real tensor and
	$\Phi_{\mu\nu} \eta^{\mu \nu}= 0$, as expected.

	From Eqs. (\ref{29}, \ref{30}, \ref{P}), we can find that $P_{ij}$ and $S_{ij}$ are also non-diagonal components of $G_{\mu \nu}$ and $T_{\mu \nu}$.
	By comparing Eq. (\ref{EQ}) with Eq. (\ref{P}), we can interpret $P_{ij}/(8\pi G)$ as a momentum and $S_{ij}/(8 \pi G)$ as a stress of a spacetime fluctuation. We can also observe from (\ref{40}) and (\ref{41}) that the component of the Riemann curvature tensor of the form $R_{j0\;pq} $ is linked to a momentum, and the form $R_{i0\;j0}$, $R_{pq\;rs}$ linked to a stress-energy.	
	
	Now we investigate $X_{ABCD}$ and $\Psi_{ABCD}$ more in detail.
Before representing $X$ and $\Psi$ in matrix form, we can check (\ref{X}) to find out whether $\Lambda=X_{AB}^{\quad AB}=R/4$ or not.
	From the properties of Riemann curvature tensor, Ricci scalar is
	\begin{eqnarray}
	R=R_{\mu \nu}^{\quad \mu \nu}=2 R_{0i}^{\quad 0i}+R_{ij}^{\quad ij} = 2 R_{0i \rho \sigma} g^{\rho 0} g^{\sigma i}+R_{ij \rho \sigma }g^{\rho i} g^{\sigma j}.
	\end{eqnarray}
	For Minckowski metric $g_{\mu\nu}=\eta_{\mu\nu}$, $R$ becomes
	\begin{eqnarray}
	R=-2R_{i0i0}+R_{ijij}.  \label{sR}
	\end{eqnarray}
	Because
	\begin{eqnarray}
	s^k_{AB} s^l_{CD} \varepsilon^{CA} \varepsilon^{DB}= \sigma_{\;A}^{k\;\;P} \varepsilon_{PB}   \sigma_{\;C}^{l\;\;Q} \varepsilon_{QD}  \varepsilon^{CA} \varepsilon^{DB} =  (\varepsilon^{CA} \sigma_{\;A}^{k\;\;D})   (\sigma_{\;C}^{l\;\;Q} \varepsilon_{QD}) \nonumber \\
	= -Tr(\bar{s}^k s^l) = \begin{pmatrix}
	-2 & (k=l)  \\
	0 & (k\neq l)
	\end{pmatrix}~,
	\end{eqnarray}
	we  can finally see that
	\begin{eqnarray}
	X_{AB}^{\quad AB}&& =X_{ABCD} \varepsilon^{CA} \varepsilon^{DB} \nonumber
	\\&& =\frac{1}{4} (-2R_{0l\;0l}+ \frac{1}{2} \epsilon_{ijl} \epsilon_{pql} R_{ij\;pq})=\frac{1}{4}(-2R_{0l\;0l}+  R_{ij\;ij})=\frac{R}{4}
		\end{eqnarray}
		from Eq. (\ref{X}).
	We have used $\epsilon_{pql}R_{0l\;pq} =0$ by Bianchi identity.

To represent the spinors $X$ and $\Psi$ in simple matrix forms, we first define
\begin{eqnarray}
&&Q_{ij} \equiv \frac{1}{2}  \epsilon_{pqj} R_{i0\;pq} +\frac{1}{2} \epsilon_{pqi}R_{j0\;pq}, \label{55}\\
&&E_{ij}\equiv R_{0i\;0j} - \frac{1}{4} \epsilon_{pqi} \epsilon_{rsj} R_{pq\;rs} , \label{56}
\\ && \Xi_{ij}\equiv R_{i0\;j0}- \frac{1}{4} \epsilon_{pqi} \epsilon_{rsj} R_{pq\;rs}-\frac{1}{2} i\epsilon_{pqj}R_{i0\;pq} -\frac{1}{2} i \epsilon_{pqi} R_{pq\;j0}  =E_{ij}-i\,Q_{ij}  , \label{xi}
\end{eqnarray}
where $Q_{ij}$ and $E_{ij}$ both are symmetric for $i,j$.
Then, we have
\begin{eqnarray}
X_{ABCD}= \frac{1}{4} \Xi_{ij}s^i_{AB} s^j_{CD} , \label{XX}
\end{eqnarray}
from (\ref{X}).
     This can be expressed in a $4 \times 4 $ matrix form by multiplying the factors in a similar way to the (\ref{P}), but here it is useful to multiply by $(\sigma \varepsilon )_{\;\mu}^{*\;\;AC} ( \sigma \varepsilon)_{\,\nu}^{*\;\;BD}=\bar{s}_\mu^{\;AC} \bar{s}_\nu^{\;BD}$ for simplicity, instead of $ \sigma_\mu^{\;AC} \sigma_\nu^{\;BD}$, where the components of $\bar{s}_\mu^{\;AC}$ is equal to $\bar{s}^\mu$ defined in (\ref{dsb}):
       \begin{eqnarray}
    &&X_{ABCD} \bar{s}_\mu^{\;AC} \bar{s}_\nu^{\;BD}=(\frac{1}{4} \Xi_{ij}s^i_{AB} s^j_{CD} ) \bar{s}_\mu^{\;AC} \bar{s}_\nu^{\;BD} \nonumber
    \\     &&=
    \frac{1}{2}\begin{pmatrix}
    -\Xi_{11}-\Xi_{22}-\Xi_{33} & -i\Xi_{23}+i\Xi_{32}& i\Xi_{13}-i\Xi_{31}&-i\Xi_{12}+i\Xi_{21}\\
    -i\Xi_{23}+i\Xi_{32}& \Xi_{11}-\Xi_{22}-\Xi_{33}& \Xi_{12}+\Xi_{21} &\Xi_{13}+\Xi_{31}\\
    i\Xi_{13}-i\Xi_{31} & \Xi_{12}+\Xi_{21}& -\Xi_{11}+\Xi_{22}-\Xi_{33}& \Xi_{23}+ \Xi_{32} \\
    -i\Xi_{12}+i\Xi_{21} & \Xi_{13}+\Xi_{31} & \Xi_{23}+\Xi_{32} & -\Xi_{11}-\Xi_{22} +\Xi_{33}
    \end{pmatrix}.
    \end{eqnarray}
   Since  $\Xi_{ij}$ is symmetric for $i,j$, it becomes
    \begin{eqnarray}
     =
     \begin{pmatrix}
    -\Xi_{11}-\Xi_{22}-\Xi_{33} & 0& 0&0\\
   0& \Xi_{11}-\Xi_{22}-\Xi_{33}& 2\Xi_{12} &2\Xi_{13}\\
   0 & 2\Xi_{12}& -\Xi_{11}+\Xi_{22}-\Xi_{33}&2\Xi_{23} \\
  0& 2\Xi_{13} & 2\Xi_{23} & -\Xi_{11}-\Xi_{22} +\Xi_{33}
    \end{pmatrix}.
    \end{eqnarray}
 For Wely conformal spinor $ \Psi_{ABCD} = \frac{1}{3}(X_{ABCD}+X_{ACDB}+X_{ADBC})$,
              \begin{eqnarray}
       &&\Psi_{ABCD} \bar{s}_\mu^{\;AC} \bar{s}_\nu^{\;BD}=\frac{1}{12} \Xi_{ij}(s^i_{AB} s^j_{CD}+s^i_{AC} s^j_{DB}+s^i_{AD} s^j_{BC} ) \bar{s}_\mu^{\;AC} \bar{s}_\nu^{\;BD}
       \nonumber  \\     &&=
       \frac{1}{3}\begin{pmatrix}
       0& -i\Xi_{23}+i\Xi_{32}& i\Xi_{13}-i\Xi_{31}&-i\Xi_{12}+i\Xi_{21}\\
     0 & 2\Xi_{11}-\Xi_{22}-\Xi_{33}& 2\Xi_{12}+\Xi_{21} &2\Xi_{13}+\Xi_{31}\\
       0& \Xi_{12}+2\Xi_{21}& -\Xi_{11}+2\Xi_{22}-\Xi_{33}& 2\Xi_{23}+\Xi_{32} \\
       0 & \Xi_{13}+2\Xi_{31} & \Xi_{23}+2\Xi_{32} & -\Xi_{11}-\Xi_{22} +2\Xi_{33} \label{63}
       \end{pmatrix}.
       \end{eqnarray}
Considering the symmetricity of $\Xi$, it becomes
  \begin{eqnarray}
=\frac{1}{3}\begin{pmatrix}
\;0\quad & 0& 0&0\\
\;0\quad& 2\Xi_{11}-\Xi_{22}-\Xi_{33}& 3\Xi_{12} &3\Xi_{13}\\
\;0\quad & 3\Xi_{12}& -\Xi_{11}+2\Xi_{22}-\Xi_{33}&3\Xi_{23} \\
\;0\quad& 3\Xi_{13} & 3\Xi_{23} & -\Xi_{11}-\Xi_{22} +2\Xi_{33}
\end{pmatrix}. \label{psi}
\end{eqnarray}%
The components of $X_{ABCD}, \Psi_{ABCD}$ are expressed as symmetric tensors. As we can see on (\ref{63}) and (\ref{psi}),  $\Xi_{ij}$ includes all information of $\Psi_{ABCD}$.
 Because of Wely tensor $C_{abcd} =\Psi_{ABCD} \varepsilon_{A'B'}\varepsilon_{C'D'} +\bar{\Psi}_{A'B'C'D'} \varepsilon_{AB}\varepsilon_{CD}$,
 we may conclude that all informations of Weyl tensor are comprehended in $\Xi_{ij}$.

The form of (\ref{psi}) is similar to the tidal tensor $\mathbb{T}_{ij}$ with a potential $U=-U_0/r=-U_0/\sqrt{x^2+y^2+z^2}$:
\begin{eqnarray}
\mathbb{T}_{ij}=\frac{U_0}{r^5} \begin{pmatrix}
2x^2-y^2-z^2 & 3xy & 3xz \\
3xy & -x^2+2 y^2-z^2 & 3yz \\
3xz & 3yz & -x^2-y^2+2 z^2
\end{pmatrix},
\end{eqnarray}
 where $\mathbb{T}_{ij}=J_{ij}-J_a^a \delta_{ij}$ and $J_{ij}=\delta^2 U/\delta x^i \delta x^j$ \cite{baldauf2012evidence,forgan2016tensor}. The similarity may come from the link between tidal forces and Weyl tensor.
 The tidal force in general relativity is described by the Riemann curvature tensor. The Riemman curvature tensor $R_{abcd}$ can be decomposed to $R_{abcd} =S_{abcd} +C_{abcd}$, where $C_{abcd}$ is a traceless part which is a Weyl tensor and $S_{abcd}$ is a remaining part which consists of Ricci tensor $R_{ab}=R^c_{\;\;acb}$ and $R=R_a^{\;a}$ \cite{carroll2004spacetime}. In the Schwartzchild metric, since $R=R_{ab}=S_{abcd}=0$ but $ C_{abcd}\neq 0$, the tidal forces are described by Weyl tensor. This shows that $C_{abcd}$, $\Psi_{ABCD}$ and $\Xi_{ij}$ are all related to the tidal effects.

The components of $\Psi$ and $\Xi$ can be represented with Weyl tensors. In a flat coordinate, by using \begin{eqnarray}
R_{\mu \rho }=R_{\mu \nu \rho \sigma }g^{\nu \sigma }=R_{\mu 0\rho 0}-R_{\mu i\rho i}
\end{eqnarray}
and Eq. (\ref{sR}), the components of Weyl tensor (\ref{C}) can be expressed as
\begin{eqnarray}
&&C_{0\underline{p}0\underline{q}}=C_{\underline{j}\,\underline{p}\,\underline{j}\,\underline{q}}=\frac{1}{2}R_{0\underline{p}0\underline{q}}+\frac{1}{2}R_{\underline{p}\underline{j}\underline{q}\underline{j}} \quad \text{  (for $p\neq q$)~, }  \label{c1}\\
&&C_{0\underline{p}0\underline{p}}=-C_{\underline{i}\,\underline{j}\,\underline{i}\,\underline{j}}=\frac{1}{2}R_{0\underline{p}0\underline{p}}-\frac{1}{2}R_{0k0k}+\frac{1}{2}R_{\underline{p}k\underline{p}k}-\frac{1}{2}R_{klkl}~,\\
&&C_{\underline{p}0\underline{p}\underline{q}}=R_{\underline{p}0\underline{p}\underline{q}}-\frac{1}{2}R_{0i\underline{q}i}~, \\
&&C_{\underline{i}0\underline{p}\underline{q}}=R_{\underline{i}0\underline{p}\underline{q}}~. \label{c2}
\end{eqnarray}
Comparing Eq. (\ref{c1}) -- Eq. (\ref{c2}) with (\ref{55}) and (\ref{56}), we find that
\begin{eqnarray}
&&C_{0\underline{p}0\underline{q}}=C_{\underline{j}\,\underline{p}\,\underline{j}\,\underline{q}}=\frac{1}{2} E_{\underline{p} \underline{q}} \quad \text{  (for $p\neq q$)~, } \\
&&C_{0\underline{p}0\underline{p}}=-C_{\underline{i}\,\underline{j}\,\underline{i}\,\underline{j}}=\frac{1}{6}(3E_{\underline{p}\underline{p}}-E_{11}-E_{22}-E_{33})~,\\
&&C_{\underline{p}0\underline{p}\underline{q}}= \epsilon_{\underline{i}\,\underline{p}\,\underline{q}}\frac{Q_{\underline{i}\,\underline{p}}}{2}~,\\
&&C_{\underline{i}0\underline{p}\underline{q}}= \epsilon_{\underline{i}\,\underline{p}\,\underline{q}} \frac{Q_{\underline{i}\underline{i}} }{2}\qquad \text{  (for $i\neq p$ and $i\neq q$) }~.
\end{eqnarray}
Therefore, (\ref{psi}) can be rewritten to
\begin{eqnarray}
\Psi_{ij} =\Psi_{ABCD} \bar{s}_i^{\;AC} \bar{s}_j^{\;BD}=2 C_{0i0j}- i\epsilon^{ipq}C_{j0pq} +\frac{i}{3}\epsilon^{lpq}C_{l0pq} ~.
\end{eqnarray}
Since $\epsilon^{lpq}C_{l0pq}=Q_{11}+Q_{22}+Q_{33}$ is zero by Bianchi identity, it becomes
\begin{eqnarray}
\Psi_{ij} =\Psi_{ABCD} \bar{s}_i^{\;AC} \bar{s}_j^{\;BD}=2 C_{0i0j}- i\epsilon^{ipq}C_{j0pq}~.
\end{eqnarray}
And Eq. (\ref{xi}) can be reformulated  to
\begin{eqnarray}
\Xi_{ij} = 2 C_{0i0j}-\frac{R}{6}\delta_{ij}-i\epsilon^{ipq}C_{j0pq}~,
\end{eqnarray}
where $R=-2E_{ii}=-2\Xi_{ii}=-2(E_{11}+E_{22}+E_{33})$.
Therefore, we can  finally find the relation
\begin{eqnarray}
\Xi_{ij} = \Psi_{ij}-\frac{R}{6}\delta_{ij}.
\end{eqnarray}
 Here we can see the equivalence and the direct correspondences among $\Psi_{ABCD}$, $\Xi_{ij}$ and Weyl tensor.

	\section{Definition of Sedon and Relations among spinors, sedenion and sedon }

	In this section, we investigate the basis of sedenion and we define a new algebra which is a similar structure to sedenion.
	Sedenion is 16 dimensional noncommutative and nonassociative algebra, which can be obtained from Cayley-Dickson construction \cite{cowles2017cayley, saniga2014cayley}.
	The multiplication table of sedenion basis is shown in Table \ref{table2}.
The elements of sedenion basis can be represented in the form $e_i= q^{\mu} \otimes q'^{\mu'}=q^{\mu\mu'}$ with  $i=\mu+4\nu$,  where
	$q^\mu=(1,\mathbf{i},\mathbf{j},\mathbf{k})$,   $q'^{\mu'}=(1,\mathbf{i}',\mathbf{j}',\mathbf{k}')$. The multiplication rule can be written by
	$e_i *e_j=q^{\mu\mu'}*q^{\nu\nu'}= s_{\mu\mu' \nu\nu' } \; q^{\mu\mu'} q^{\nu \nu'}$, where $s_{\mu\mu' \nu\nu' }$ is +1 or -1, which is determined by $\mu,\mu', \nu,\nu'$ \cite{hong2019quaternion}.

\begin{table}	
\begin{tabular}{|c||*{4}{c}|*{4}{c}|*{4}{c}|*{4}{c}|}\hline
	&e0	&e1	&e2	&e3	&e4	&e5	&e6	&e7	&e8	&e9	&e10	&e11	&e12	&e13	&e14	&e15\\ \hline \hline
	e0	&e0	&e1	&e2	&e3	&e4	&e5	&e6	&e7	&e8	&e9	&e10	&e11	&e12	&e13	&e14	&e15\\
	e1	&e1	&-e0	&e3	&-e2	&e5	&-e4	&-e7	&e6	&e9	&-e8	&-e11	&e10	&-e13	&e12	&e15	&-e14\\
	e2	&e2	&-e3	&-e0	&e1	&e6	&e7	&-e4	&-e5	&e10	&e11	&-e8	&-e9	&-e14	&-e15	&e12	&e13\\
	e3	&e3	&e2	&-e1	&-e0	&e7	&-e6	&e5	&-e4	&e11	&-e10	&e9	&-e8	&-e15	&e14	&-e13	&e12\\ \hline
	e4	&e4	&-e5	&-e6	&-e7	&-e0	&e1	&e2	&e3	&e12	&e13	&e14	&e15	&-e8	&-e9	&-e10	&-e11\\
	e5	&e5	&e4	&-e7	&e6	&-e1	&-e0	&-e3	&e2	&e13	&-e12	&e15	&-e14	&e9	&-e8	&e11	&-e10\\
	e6	&e6	&e7	&e4	&-e5	&-e2	&e3	&-e0	&-e1	&e14	&-e15	&-e12	&e13	&e10	&-e11	&-e8	&e9\\
	e7	&e7	&-e6	&e5	&e4	&-e3	&-e2	&e1	&-e0	&e15	&e14	&-e13	&-e12	&e11	&e10	&-e9	&-e8\\ \hline
	e8	&e8	&-e9	&-e10	&-e11	&-e12	&-e13	&-e14	&-e15	&-e0	&e1	&e2	&e3	&e4	&e5	&e6	&e7\\
	e9	&e9	&e8	&-e11	&e10	&-e13	&e12	&e15	&-e14	&-e1	&-e0	&-e3	&e2	&-e5	&e4	&e7	&-e6\\
	e10	&e10	&e11	&e8	&-e9	&-e14	&-e15	&e12	&e13	&-e2	&e3	&-e0	&-e1	&-e6	&-e7	&e4	&e5\\
	e11	&e11	&-e10	&e9	&e8	&-e15	&e14	&-e13	&e12	&-e3	&-e2	&e1	&-e0	&-e7	&e6	&-e5	&e4\\ \hline
	e12	&e12	&e13	&e14	&e15	&e8	&-e9	&-e10	&-e11	&-e4	&e5	&e6	&e7	&-e0	&-e1	&-e2	&-e3\\
	e13	&e13	&-e12	&e15	&-e14	&e9	&e8	&e11	&-e10	&-e5	&-e4	&e7	&-e6	&e1	&-e0	&e3	&-e2\\
	e14	&e14	&-e15	&-e12	&e13	&e10	&-e11	&e8	&e9	&-e6	&-e7	&-e4	&e5	&e2	&-e3	&-e0	&e1\\
	e15	&e15	&e14	&-e13	&-e12	&e11	&e10	&-e9	&e8	&-e7	&e6	&-e5	&-e4	&e3	&e2	&-e1	&-e0 \\ \hline
\end{tabular}
	\caption{
	The multiplication table of sedenion.  For convenience, `$e_N$'s are represented as `eN' ; e.g. $e_3 \rightarrow$ e3. }
\label{table2}
\end{table}

\begin{table}[]
	\begin{tabular}{|c||*{4}{c}|*{4}{c}|*{4}{c}|*{4}{c}|} 
		\hline
		&e0	&e1	&e2	&e3	&e4	&e5	&e6	&e7	&e8	&e9	&e10	&e11	&e12	&e13	&e14	&e15\\ \hline \hline
	e0	&e0	&e1	&e2	&e3	&e4	&e5	&e6	&e7	&e8	&e9	&e10	&e11	&e12	&e13	&e14	&e15\\
	e1	&e1	&-e0	&e3	&-e2	&e5	&-e4	&\rd{e7}	&\rd{-e6}	&e9	&-e8	&\rd{e11}&\rd{-e10}&\rd{e13}&\rd{-e12}&e15	&-e14\\
	e2	&e2	&-e3	&-e0	&e1	&e6	&\rd{-e7}	&-e4	&\rd{e5}	&e10	&\rd{-e11}&-e8	&\rd{e9}	&\rd{e14}&-e15	&\rd{-e12}&e13\\
	e3	&e3	&e2	&-e1	&-e0	&e7	&\rd{e6	}&\rd{-e5}	&-e4	&e11	&\rd{e10}&\rd{-e9}&-e8	&\rd{e15}&e14	&-e13	&\rd{-e12}\\  \hline
	e4	&e4	&\rd{e5}	&\rd{e6}	&\rd{e7}	&-e0	&\rd{-e1	}&\rd{-e2}	&\rd{-e3}	&e12	&e13	&e14	&e15	&-e8	&-e9	&-e10	&-e11\\
	e5	&e5	&\rd{-e4}	&\rd{e7	}&\rd{-e6}	&-e1	&\rd{e0}	&-e3	&e2	&e13	&-e12	&e15	&-e14	&\rd{-e9}	&\rd{e8}	&\rd{-e11}&\rd{e10}\\
	e6	&e6	&\rd{-e7}	&\rd{-e4	}&\rd{e5}	&-e2	&e3	&\rd{e0}	&-e1	&e14	&-e15	&-e12	&e13	&\rd{-e10}&\rd{e11}&\rd{e8}&\rd{-e9}\\
	e7	&e7	&\rd{e6}	&\rd{-e5}	&\rd{-e4	}&-e3	&-e2	&e1	&\rd{e0}	&e15	&e14	&-e13	&-e12	&\rd{-e11}&\rd{-e10}&\rd{e9}&\rd{e8}\\  \hline
	e8	&e8	&\rd{e9}	&\rd{e10}&\rd{e11}&-e12	&-e13	&-e14	&-e15	&-e0	&\rd{-e1}	&\rd{-e2	}&\rd{-e3}	&e4	&e5	&e6	&e7\\
	e9	&e9	&\rd{-e8}	&\rd{e11}&\rd{-e10}&-e13	&e12	&\rd{-e15}&\rd{e14}&-e1	&\rd{e0}	&-e3	&e2	&\rd{e5}&\rd{-e4}	&e7	&-e6\\
	e10	&e10	&\rd{-e11}&\rd{-e8}&\rd{e9}	&-e14	&\rd{e15}&e12	&\rd{-e13}&-e2	&e3	&\rd{e0}	&-e1	&\rd{e6	}&-e7	&\rd{-e4	}&e5\\
	e11	&e11	&\rd{e10}&\rd{-e9}&\rd{-e8}	&-e15	&\rd{-e14}&\rd{e13}&e12	&-e3	&-e2	&e1	&\rd{e0}	&\rd{e7}	&e6	&-e5	&\rd{-e4}\\  \hline
	e12	&e12	&e13	&e14	&e15	&e8	&\rd{e9}	&\rd{e10}&\rd{e11}&-e4	&\rd{-e5	}&\rd{-e6}	&\rd{-e7}	&-e0	&-e1	&-e2	&-e3\\
	e13	&e13	&-e12	&e15	&-e14	&e9	&\rd{-e8	}&e11	&-e10	&-e5	&\rd{e4}	&\rd{-e7}	&\rd{e6	}&\rd{-e1}	&\rd{e0}	&\rd{-e3}	&\rd{e2}\\
	e14	&e14	&-e15	&-e12	&e13	&e10	&-e11	&\rd{-e8}	&e9	&-e6	&\rd{e7	}&\rd{e4}	&\rd{-e5}	&\rd{-e2}	&\rd{e3}	&\rd{e0}	&\rd{-e1}\\
	e15	&e15	&e14	&-e13	&-e12	&e11	&e10	&-e9	&\rd{-e8}	&-e7	&\rd{-e6}	&\rd{e5}	&\rd{e4	}&\rd{-e3}	&\rd{-e2}	&\rd{e1}	&\rd{e0} \\ \hline
	\end{tabular}
	\caption{
		The multiplication table of sedon.}
	\label{table3}
\end{table}

	\begin{table}[]
		\begin{tabular}{|c||c|c|c|c|}
			\hline
			& $\; \sim \otimes \; \mathbf{1}\;$ &$\; \sim \otimes \; \mathbf{i}\;$ &$\; \sim \otimes \; \mathbf{j}\;$&$\; \sim \otimes \; \mathbf{k}\;$\\ \hline \hline
		$\; \mathbf{1} \; \otimes  \sim \;$ & $A_0$ & $B_1$  & $B_1$ & $B_1$ \\ \hline
		$\; \mathbf{i} \; \otimes  \sim \;$ & $C_1$ & $D_{11}$  & $D_{12}$ & $D_{13}$ \\ \hline
		$\; \mathbf{j} \; \otimes  \sim \;$ & $C_2$ & $D_{21}$  & $D_{22}$ & $D_{23}$ \\ \hline
		$\; \mathbf{k} \; \otimes  \sim \;$ & $C_3$ & $D_{31}$  & $D_{32}$ & $D_{33}$ \\ \hline
		\end{tabular}
	\caption{
		The representation of coefficients of sedon.
	} \label{table4}
	\end{table}

	Table \ref{table3}  shows the multiplication table of an algebra which is similar to sedenion. It is consisted of 16 basis $e_i=q^{\mu} \otimes q'^{\mu'} $  with  $i=\mu+4\nu$ and the multiplication rule $e_i *e_j=(q^{\mu} \otimes q'^{\mu'} ) * (q^{\mu} \otimes q'^{\mu'})= (q^{\mu} q^{\mu} \otimes q'^{\mu'} q'^{\mu'})$. The table is almost the same as the multiplication table of sedenion basis, but just differs in signs. The signs of red colored elements in Table \ref{table3} differ from Table \ref{table2}. We will call this algebra as `sedon'.
	
	Sedon can be written in the form
	\begin{eqnarray}
	S= A_0 +| \vec{B} \} +\{ \vec{C} |+ \{\overleftrightarrow{D}\}=A_0 + B_i q_R^i + C_i q_L^i + D_{ij} u^{ij},
	\end{eqnarray}
	where $q_R^i=1 \otimes q^i$, $q_L^i=q^i \otimes 1$, $u^{ij}= q^i \otimes q^j$, $| \vec{B} \}=B_i q_R^i $, $\{ \vec{C}| =C_i q_L^i $, and $\{\overleftrightarrow{D}\}=D_{ij} u^{ij}$.
	We can name $| \vec{B} \}$ as `right svector', $\{ \vec{C}|$ as `left svector', and $\{\overleftrightarrow{D}\}$ as `stensor'. The coefficient of sedon can be represented as in Table \ref{table4}. For example, $D_{13}$ is a coefficient of $i \otimes k $ term.
	
	 Now we will see the relation between Ricci spinors and the sedon. Since
	\begin{eqnarray}
	\sigma^{i\;\;C}_A \varepsilon_{CB}=s^i_{AB}, \qquad \varepsilon_{A'C'} \sigma^{C'}_{\;\;B'}=-\bar{s}^i_{A'B'},
	\end{eqnarray}
	therefore
	\begin{eqnarray}
	\sigma_{A}^{i\;\;B} = -s^i_{AC} \varepsilon^{CB},  \qquad \sigma^{i\;C'}_{\;\;\;\;B'}=\varepsilon^{C'A'}\bar{s}^i_{A'B'}.
	\end{eqnarray}
	Eq. (\ref{phi}) can be reformulated as
	\begin{eqnarray}
	\Phi_{A\quad D'}^{\;\;\;QP'}= \varepsilon^{P'C'} \Phi_{ABC'D'} \varepsilon^{BQ} =   \frac{1}{4} \Theta_{ij} \varepsilon^{P'C'} s^i_{AB}\bar{s}^j_{C'D'} \varepsilon^{BQ} = - \frac{1}{4} \Theta_{ij} \sigma^{i\;\;\;Q}_{\;A} \sigma^{j \,P'}_{\;\;\;\;D'}  . \label{pphi}
	\end{eqnarray}
	Since $q^i=(\mathbf{i},\mathbf{j},\mathbf{k})$ is isomorphic to $-i\sigma^i = (-\sigma^1 i,-\sigma^2 i,- \sigma^3 i)$, we can set $q^i=-i\sigma^i$.
	And Eq. (\ref{pphi}) can be written as
	\begin{eqnarray}
	\Phi_{A\quad D'}^{\;\;\;QP'} = \frac{1}{4} \Theta_{ij} q^{i\;\;\;Q}_{\;A} q^{j \,P'}_{\;\;\;\;D'}~, \label{Q}
	\end{eqnarray}
	which can be regarded as a sedon.
	In a similar way, $X_{ABCD}$ can be written as
	\begin{eqnarray}
	X_{A\;\;C}^{\;\;B\;\;D}=\frac{1}{4}\Xi_{kl}\, q^{k\;\;\;B}_{\;A} q^{l\;\;\;C}_{\;D}.
	\end{eqnarray}
	From Eq. (\ref{Q}), a Ricci spinor can be interpreted as a combination of a right-handed and a left-handed rotational operations, since the basis has the form `left-handed quaternion $\otimes$ right- handed quaternion'. Following the rotational interpretation of Cayley-Dickson algebra \cite{hong2019quaternion}, it can be interpreted as the 2-fold rotation $\otimes$ 2-fold rotation.

		For two quaternions $\d{A}=A_i q^i= a_1 \mathbf{i} + a_2 \mathbf{j} + a_3  \mathbf{k}$ and $\d{B}=B_j q^j=b_1 \mathbf{i} + b_2 \mathbf{j} + b_3  \mathbf{k}$, which can be represented in the $2 \times 2$ matrix representation with spinor indices ($A_i q^{i \;\; D}_{\;C}$ and $B_j q^{j \;\; D}_{\;C}$),  the multiplication of them can be written as
	\begin{eqnarray}
	\d{A}\d{B}=A_i q^{i \;\; D}_{\;C} \;  B_j q^{j \;\; E}_{\;D}= -A_iB_i \delta_{C}^{\;\;E} +\epsilon_{ijk}A_i B_j q^{k\;\;E}_{\;C}. \label{qr}
	\end{eqnarray}
	We can use this to express multiplications of spinors.  One of the example is
	\begin{eqnarray}
	\Phi_{A\quad D'}^{\;\;\;BC'}\Phi_{B\quad F'}^{\;\;\;ED'}
	&&=\theta_{ij} q^{i \;\; B}_{\;A}q^{j \,C'}_{\;\;\;\;D'} \; \theta_{rs} q^{r \;\; E}_{\;B}q^{s \,D'}_{\;\;\;\;F'}
	\nonumber\\
	&&=(-\theta_{lj}\theta_{ls} \delta_A^{\;\;\;E}+\epsilon_{pqu} \theta_{pj} \theta_{qs} q^{u \;\; E}_{\;A})q^{j \,C'}_{\;\;\;\;D'}q^{s \,D'}_{\;\;\;\;F'}
	\nonumber\\
	&&=\theta_{lk} \theta_{lk}\;\; \delta_A^{\;\;\;E}\delta_{\;\;\;F'}^{C'}-\epsilon_{mnv}\theta_{lm} \theta_{ln} \;\; q^{v \,C'}_{\;\;\;\;F'}\delta_A^{\;\;\;E}
	\nonumber\\&&\;\;\;-\epsilon_{pqu}\theta_{pl} \theta_{ql} \;\;  q^{u \;\; E}_{\;A} \delta_{\;\;\;F'}^{C'}+\epsilon_{mnv}\epsilon_{pqu}  \theta_{pm}\theta_{qn} \;\; q^{u \;\; E}_{\;A} q^{v \,C'}_{\;\;\;\;F'},
	\end{eqnarray}
	where $\theta_{ij}=\frac{1}{4} \Theta_{ij}$. 	The result is also a sedon form. Above example shows not only multiplications of $\Phi_{A\quad D'}^{\;\;\;BC'}$
	 but also the general multiplication of stensor.
	Here is an another example:
	An antisymmetric differential operator $\nabla_{[a}\nabla_{b]}$ can be divided into two parts
	\begin{eqnarray}
	\Delta_{ab}=2\nabla_{[a}\nabla_{b]}=\epsilon_{A'B'} \square_{AB} +\epsilon_{AB} \square_{A'B'},
	\end{eqnarray}
	where $\square_{AB}=\frac{1}{2} \Delta_{AA'B}^{\qquad A'}$
	and $\square_{A'B'}=\frac{1}{2} \Delta_{AA'\;\;B'}^{\quad \, A}$.
	As we can see in (\ref{86}) and (\ref{88}), each term can be considered as a quaternion.	
	\begin{eqnarray}
	&&\square_A^{\;\;\;B}=\frac{1}{4} \Delta_{\mu\nu}\sigma^\mu_{AA'}\bar{\sigma}^{\nu\,A'B}=\frac{1}{2}(i\,\Delta_{k0} +\frac{1}{2}\epsilon_{\;\;k}^{ij} \Delta_{ij})\;q^{k \;\; B}_{\;A}
	\\
		&&\bar{\square}^{A'}_{\;\;B'}=-\frac{1}{4}\Delta_{\mu\nu} \;\bar{\sigma}^{\mu A'C} \sigma^{\nu}_{CB'}
	= \frac{1}{2}(i\,\Delta_{k0}-\frac{1}{2}\, \epsilon_{\;\;k}^{ij} \Delta_{ij} ) \; q^{k \,A'}_{\;\;\;\;B'}.	\end{eqnarray}
	Then, $\square_A^{\;\;\;B} \Phi_{B\quad E'}^{\;\;\;CD'}$ can be considered as a multiplication of a quaternion and a sedon.
	\begin{eqnarray}
	\square_A^{\;\;\;B} \Phi_{B\quad E'}^{\;\;\;CD'}&&=\flat_k\; q^{k \;\; B}_{\;A}\theta_{ij}q^{i \;\; C}_{\;B}q^{j \,D'}_{\;\;\;\;E'}\nonumber\\&&=-\flat_p \;\theta_{pj}\delta_A^{\;\;\;C}q^{j \,D'}_{\;\;\;\;E'}+\epsilon_{kip}\flat_k\;\theta_{ij} q^{p \;\; C}_{\;A}q^{j \,P}_{\;\;\;\;T}\nonumber\\
	 &&= -i\, \Delta_{p0} \theta_{pj} \;\; \delta_A^{\;\;\;C} q^{j \,D'}_{\;\;\;\;E'}-\frac{1}{2}\epsilon_{rsp}\Delta_{rs}\theta_{pj} \;\; \delta_A^{\;\;\;C} q^{j \,D'}_{\;\;\;\;E'} \nonumber\\&& \;\; \; +i \, \epsilon_{kip}\Delta_{k0} \theta_{ij} \;\; q^{p \;\; C}_{\;A}q^{j \,D'}_{\;\;\;\;E'} +\frac{1}{2}\epsilon_{lqk}\Delta_{lq}\epsilon_{kip}\theta_{ij} \;\; q^{p \;\; C}_{\;A}q^{j \,D'}_{\;\;\;\;E'} , \label{62}
	 \end{eqnarray}
	 where $\flat_k \equiv i\Delta_{k0} +\frac{1}{2}\epsilon_{ijk} \Delta_{ij}$.
	$\epsilon_{lqk}\Delta_{lq}\epsilon_{kip}\theta_{ij}$ in the last term can be changed as $\Delta_{qp}\theta_{qj}-\Delta_{pq}\theta_{qj}=2 \Delta_{qp} \theta_{qj}$. The result (\ref{62}) is in a sedon form.
	Using those expressions, we can represent the quantities with spinor indices as sedon forms whose elements are components of tensors.

		\section{A few Examples of Curvature Spinors in a locally flat coordinate}

	\subsection{Weyl conformal spinor for the Schwarzschild metric: An Example of Section IV \label{Ex1}}

	It is known that the Schwarzschild metric can be represented in Fermi normal coordinate \cite{manasse1963fermi}.
		In Schwarzschild coordinate $x^{\mu'}=(T,R,\Theta, \Phi)$, the metric is displayed in the form
		\begin{eqnarray}
		ds^2=g_{\mu' \nu'}dy^{\mu'} dy^{\nu'}=-f dT^2+ f^{-1} dR^2+R^2 d\Theta^2 +R^2 \rm{sin}^2 \Theta \; d\Phi^2~,
		\end{eqnarray}
		where $f=1-2GM/R$.
		The basis of a constructed Fermi coordinate $x^\mu =(t,x,y,z)$ is
		\begin{eqnarray}
		\mathbf{e}_0 &&=\partial /\partial t |_G = T' \; \partial /\partial T+R' \; \partial /\partial R,\nonumber \\
		\mathbf{e}_1 &&=\partial /\partial x |_G =X^{-1} R' \; \partial /\partial T+fT' \;\partial /\partial R,\nonumber \\
		\mathbf{e}_2 &&=\partial /\partial y |_G = 1/R \;\partial /\partial \Theta,\nonumber \\
		\mathbf{e}_3 &&=\partial /\partial z |_G = 1/R\; \rm{sin} \Theta \; \partial /\partial \Phi,
		\end{eqnarray}
		where the primes indicate derivatives with respect to proper time $t$.
		The non-zero components of the Riemann curvature tensor $R_{\mu' \nu' \rho' \sigma'}$ in Schwarzschild coordinate are
		\begin{eqnarray}
		R_{1'0'1'0'}&&=2GM/R^2 ,\nonumber \\
		R_{3'0'3'0'}&&=-(f GM/R) \rm{sin}^2 \Theta ,\nonumber \\
		R_{1'2'1'2'}&&=GM/(f R) ,\nonumber\\
		R_{2'0'2'0'}&&=-f GM/R ,\nonumber\\
		R_{2'3'2'3'}&&=-2GM R \rm{sin}^2 \Theta ,\nonumber\\
		R_{1'3'1'3'}&&=(GM/RX) \rm{sin}^2 \Theta  .
		\end{eqnarray}
		Then the Riemman curvature tenor $R_{\mu \nu \rho \sigma}$ in the Fermi coordinate is
		\begin{eqnarray}
		R_{10\;10}&&=2GM/R^3, \nonumber\\
		R_{20\;20}&&=R_{30\;30}=-GM/R^3, \nonumber\\
		R_{12\;12}&&=R_{13\;13}=GM/R^3 ,\nonumber\\
		R_{23\;23}&&=-2GM/R^3.
		\end{eqnarray}
		From Eqs. (\ref{40}), (\ref{41}), (\ref{55}) and (\ref{56}),
		we can observe that $P_{ij}=S_{ij}=Q_{ij}=0$, but $ E_{11}=4GM/R^3$ and $E_{22}=E_{33}=-2GM/R^3$. Classically, the tidal acceleration of black hole along the radial line is $-2GM/r^3 \delta X$, and the acceleration perpendicular to the radial line is $GM/r^3 \delta X$, where $\delta X$ is the separation distance of two test particles.
		In this example, the link between $\Xi_{ij}$ and tidal accelerations has been shown.

 \subsection{The spinor form of the Einstein-Maxwell equation: An Example of Section V}

		Einstein-Maxwell Equations, which is Einstein Equations in presence of electromagnetic fields, is known \cite{santos2016introduction} as
		\begin{eqnarray}
		R_{\mu\nu}-\frac{1}{2}g_{\mu\nu}R= 8 \pi G (F_{\mu\sigma}F_\nu^\sigma -g_{\mu\nu} \frac{1}{4}F_{\rho \sigma}F^{\rho \sigma} ),
		\end{eqnarray}
		where $F_{\mu\sigma}F_\nu^\sigma -g_{\mu\nu} \frac{1}{4}F_{\rho \sigma}F^{\rho \sigma} $ is the electromagnetic stress-energy tensor.
		The spinor form of the Einstein-Maxwell equation \cite{penrose1984spinors} is
		\begin{eqnarray}
		\Phi_{ABA'B'}=8 \pi G\varphi_{AB}\bar{\varphi}_{A'B'},
		\end{eqnarray}
		where  $\varphi_{AB}, \bar{\varphi}_{A'B'}$ are decomposed spinors of electromagnetic tensor $F_{\mu \nu}$, as following  (\ref{p1}) and (\ref{p2}).	
	This can be deformed to
	\begin{eqnarray}
	\Phi_{A\quad B'}^{\;\;\;BA'}=8 \pi G\varphi_A^{\;\;B}\bar{\varphi}^{A'}_{\;\;B'}.
	\end{eqnarray}
	From Eq. (\ref{P}),
	\begin{eqnarray}
	&&	 - \frac{1}{4} \Theta_{kl} \sigma^{k\;\;\;B}_{\;A} \sigma^{l \,A'}_{\;\;\;\;B'} \nonumber\\
	&&=8 \pi G[\frac{1}{2}(F_{k0}-\frac{1}{2}i\,\epsilon_{\;\;k}^{ij}F_{ij}) \sigma^{k\;\;\;B}_{\;A} ]\times
	[\frac{1}{2}(F_{l0}+\frac{1}{2}i\, \epsilon_{\;\;\;l}^{pq} F_{pq}) \sigma^{l \,A'}_{\;\;\;\;B'}]
	\nonumber \\&&=2 \pi G[ (F_{k0}F_{l0}+\frac{1}{4}\epsilon_{\;\;k}^{ij}\epsilon_{\;\;\;l}^{pq}F_{ij}F_{pq}) + \frac{i}{2}(F_{k0} \epsilon_{\;\;\;l}^{pq}F_{pq} -F_{l0}\epsilon_{\;\;k}^{ij}F_{ij})] \sigma^{k\;\;\;B}_{\;A} \sigma^{l \,A'}_{\;\;\;\;B'}~. \label{89}
	\end{eqnarray}
	Comparing the first line with the third line in (\ref{89}), we get
	\begin{eqnarray}
	S_{kl}&&= -8\pi G(F_{k0}F_{l0}+\frac{1}{4}\epsilon_{\;\;k}^{ij}\epsilon_{\;\;\;l}^{pq}F_{ij}F_{pq}), \label{fs}\\
	P_{kl}&&=-4 \pi G(F_{k0} \epsilon_{\;\;\;l}^{pq}F_{pq} -F_{l0}\epsilon_{\;\;k}^{ij}F_{ij}),
	\label{fp}
	\end{eqnarray}
	and, from Eq. (\ref{fp}) we get
	\begin{eqnarray}
	\epsilon^{mkl} P_{kl}=- 16 \pi GF_{k0}F^{mk}.
	\end{eqnarray}
	For $F^{\mu\nu}$ such that
	\begin{eqnarray}
	F^{\mu \nu}=\begin{pmatrix}
	0 & -E_1 & -E_2 & -E_3 \\
	E_1 & 0 & -B_3 & B_2 \\
	E_2 & B_3 &  0  & -B_1 \\
	E_3 & -B_2 & B_1 & 0  &
	\end{pmatrix},
	\end{eqnarray}
we have	$F_{k0}F^{mk}=(\vec{E} \times \vec{B})^m$ and $(F_{k0}F_{l0}+\frac{1}{4}\epsilon_{\;\;k}^{ij}\epsilon_{\;\;\;l}^{pq}F_{ij}F_{pq})=E_k E_l +B_k B_l$.
	From Eqs. (\ref{ep}) and (\ref{fp}) we get
	\begin{eqnarray}
	-\frac{1}{2}\epsilon^{mkl} P_{kl}&&=R_{0i}^{\;\;\;\; mi}=8\pi G(\vec{E} \times \vec{B})^k~.
	\end{eqnarray}
	This is a momentum of electromagnetic tensor and it shows that $P_{ij}/(8\pi G)$ is related to momentum.
	From (\ref{S}) and (\ref{fs}),
	\begin{eqnarray}
	S_{kl}&&=R_{0i\;0j} + \frac{1}{4} \epsilon_{pqi} \epsilon_{rsj} R_{pq\;rs}=8 \pi G(-E_k E_l -B_k B_l),\\
	S_{l\,l}&& = 8 \pi G (|\vec{E}|^2 +|\vec{B}|^2).
	\end{eqnarray}
	Those are  the shear stress and the energy of electromagnetic field.
It shows that $S_{ij}/(8 \pi G)$ is related to stress-energy.
	
	 \subsection{The quaternion form of differential Bianchi identity: Another Example  of Section V}

	The spinor form of Bianchi identity (\ref{B2}) is known \cite{penrose1984spinors} as
	\begin{eqnarray}
	\nabla^A_{B'} X_{ABCD}=\nabla^{A'}_B \Phi_{CDA'B'},
	\end{eqnarray}
	which can be deformed to
	\begin{eqnarray}
	\nabla^{B'A} X_{A\;\;C}^{\;\;B\;\;D}=\nabla^{BA'} \Phi_{C\;\;A'}^{\;\;D\;\;B'}. \label{Bii}
	\end{eqnarray}
	In flat coordinate, $\nabla^{A'A}$ equals to
	\begin{eqnarray}
	\partial^{A'A}=g^{\mu\;A'A}\partial_\mu=\frac{1}{\sqrt{2}} \sigma^{\mu A'A} \partial_\mu=\frac{1}{\sqrt{2}} \sigma^{\tilde{\mu} A'A} \partial_{\tilde{\mu}} \nonumber\\
	= \frac{1}{\sqrt{2}} \bar{q}^{\mu\; A'A} \tilde{\partial}_\mu=\frac{1}{\sqrt{2}} q^{\mu\; A'A} \partial'_\mu~,
	\end{eqnarray}
	where $\tilde{\mu}$ is tilde-spacetime indices which is defined as
	$O^{\tilde{\mu}}= (O^0,i \,O^1 ,i \,O^2 ,i \,O^3 )$, $O_{\tilde{\mu}}= (O_0,-i \,O_1 ,-i \,O_2,-i \,O_3)$ for any $O^{\mu}=(O^0,O^1,O^2,O^3)$,
	$O_{\mu}=(O_0,O_1,O_2,O_3)$ \cite{hong2019quaternion}; $\bar{q}^\mu$ is $\bar{q}^\mu=\sigma^{\tilde{\mu} A'A}=(\sigma^0,i \sigma^1,i \sigma^2, i\sigma^3)$ which is isomorphic to $(1,-\mathbf{i},-\mathbf{j},-\mathbf{k})$,  $\tilde{\partial}_\mu=\partial_{\tilde{\mu}}=(\partial_0,-i\partial_1,-i\partial_2,-i\partial_3)$, and $\partial'_\mu=(\partial_0,i\partial_1,i\partial_2,i\partial_3)$. We used the property $A_\mu B^\mu=A_{\tilde{\mu}} B^{\tilde{\mu}}$ \cite{hong2019quaternion}.  $\partial^{A'A}$ can be expanded to
	\begin{eqnarray}
	\partial^{A'A}&&=\partial_0 \delta^{A'A} +\partial'_k q^{k\; A'A}
	\end{eqnarray}
	and, considering matrix representation, $\partial^{AA'}$ can be represented as
	\begin{eqnarray}
	\partial^{AA'}&&=\partial_0 \delta^{AA'} +\partial'_k q^{\bar{k}\; AA'}=\partial_0 \delta^{AA'} + \partial'_{\bar{k}} q^{k\; AA'}~,
	\end{eqnarray}
	where the bar index $A^{\bar{k}}$ means the opposite-handed quantity of $A^k$, which is $A^{\bar{1}}=A^1,A^{\bar{2}}=-A^2,A^{\bar{3}}=A^3$ ; when $k=2$, $\bar{k}$ index change sings of $A^k$.
	It has following properties,
	\begin{eqnarray}
	A^{\bar{k}} B_k = A^k B_{\bar{k}}, \quad A^{\bar{k}}B_{\bar{k}}=A^kB_k, \quad \varepsilon_{pqr}A^{\bar{q}} B^{\bar{r}}= -\varepsilon_{\bar{p}qr}A^q B^r,\quad \varepsilon_{\bar{p}\bar{q}\bar{r}}=-\varepsilon_{pqr}.
	\end{eqnarray}
	Then Eq. (\ref{Bii}) can be written as
	\begin{eqnarray}
	(\partial_0 \delta^{B'A} +\partial'_k q^{k\; B'A})\, \Xi_{ir} q^{i\;\;B}_{\;A} q^{r\;\;D}_{\;C}= (\partial_0 \delta^{BA'} +\partial'_{\bar{k}} q^{k\; BA'})\, \Phi_{r\bar{s}}q^{r\;\;D}_{\;C} q^{s\;\;B'}_{\;A'},
	\end{eqnarray}
	since $q^{\bar{s}\;\;B'}_{\;A'}=\varepsilon^{B'D'} \varepsilon_{C'A'} q^{s \,C'}_{\;\;\;\;D'}$. Using Eq. (\ref{qr}),
	\begin{eqnarray}
	&&\partial_0 \Xi_{ir}  q^{i\;B'B} +\partial'_k \Xi_{kr} \delta^{B'B}+ \varepsilon_{pki}\partial'_k \Xi_{ir} q^{p\; B'B} \nonumber \\ &&= 	\partial_0 \Phi_{rs}  q^{\bar{s}\;BB'} +\partial'_k \Phi_{rk} \delta^{BB'}+ \varepsilon_{pks}\partial'_{\bar{k}} \Phi_{r\bar{s}} q^{p\; B'B}~,
	\end{eqnarray}
	which can be rearranged as
	\begin{eqnarray}
	\partial'_k (\Xi_{kr}-\Phi_{rk}) \delta^{B'B}+[\partial_0 (\Xi_{sr}-\Phi_{rs}) +\varepsilon_{ski}\partial'_k (\Xi_{ir}+\Phi_{ri})] q^{s\;B'B} =0~.
	\end{eqnarray}
	This is the quaternion form of Bianchi identities.

	\section{Conclusion}

		We established a new method to express curvature spinors, which allows us to grasp components of the spinors easily in a locally inertial frame.
		During such a process, we technically utilized modified sigma matrices as a basis, which are sigma matrices multiplied by $\varepsilon$, and calculated the product of sigma matrices with mixed spinor indices. Using those modified sigma matrices as a basis can be regarded as the rotation of the basis of four sigma matrices  $(\sigma^0,\sigma^1,\sigma^2,\sigma^3)$ to $(s^0,s^1,s^2,s^3)$ defined in Eq. (\ref{ds}), similar to a rotation of quaternion basis as shown in our previous work \cite{hong2019quaternion}.  By comparing the Ricci spinor with the spinor form of Einstein equation, we could appreciate the roles of each component of the Riemann curvature tensor. Furthermore, from the representation of Weyl conformal spinor, we find that the components of Weyl tensor can be replaced by complex quantities $\Xi_{ij}$, which are defined in Eq. (\ref{xi}).		
		
		We represented the elements of sedenion basis as the direct product of elements of quaternion basis themselves. And then we defined a new algebra `sedon', which has the same basis representation except slightly modified multiplication rule from the multiplication rule of sedenion. The relations between sedon and the curvature spinors are derived for a general gravitaional field, not just for a weak gravitational field. We calculated multiplications of spinors with a quaternion form, 
and observed that the results of the multiplications are also represented in a sedon form.
	
		We observed that many gravitational quantities can be represented with 3-dimensional basis. It shows that time and space may be interpreted differently from conventional interpretations in which time and space are treated as the same. And the relations among quaternion, sedon and the representations of curvature spinors imply that gravity may come from a combination of  right-handed and  left-handed {\it abstract} rotational operations.

	\acknowledgments
	This work was supported by the National Research Foundation of Korea
	(NRF) grant funded by the Korean government (MSIP) (NRF-2018R1A4A1025334).
	\\

	\nocite{*}
	\bibliographystyle{ieeetr}
	\bibliography{bibb}

\begin{thebibliography}{10}

\bibitem{penrose1984spinors}
R.~Penrose and W.~Rindler, ``Spinors and spacetime vol. 1 cambridge univ,''
  1984.

\bibitem{bain2000coordinate}
J.~Bain, ``The coordinate-independent 2-component spinor formalism and the
  conventionality of simultaneity,'' {\em Studies in History and Philosophy of
  Science Part B: Studies in History and Philosophy of Modern Physics},
  vol.~31, no.~2, pp.~201--226, 2000.

\bibitem{carmeli2000theory}
M.~Carmeli and S.~Malin, {\em Theory of spinors: An introduction}.
\newblock World Scientific Publishing Company, 2000.

\bibitem{penrose1960spinor}
R.~Penrose, ``A spinor approach to general relativity,'' {\em Annals of
  Physics}, vol.~10, no.~2, pp.~171--201, 1960.

\bibitem{o2003introduction}
J.~O'donnell~Peter, {\em Introduction to 2-spinors in general relativity}.
\newblock World Scientific, 2003.

\bibitem{carroll2004spacetime}
S.~M. Carroll, {\em Spacetime and geometry. An introduction to general
  relativity}.
\newblock 2004.

\bibitem{hong2019quaternion}
I.~K. Hong and C.~S. Kim, ``Quaternion electromagnetism and the relation with
  two-spinor formalism,'' {\em Universe}, vol.~5, no.~6, p.~135, 2019.

\bibitem{klein2008general}
D.~Klein and P.~Collas, ``General transformation formulas for fermi--walker
  coordinates,'' {\em Classical and Quantum Gravity}, vol.~25, no.~14,
  p.~145019, 2008.

\bibitem{marzlin1994physical}
K.-P. Marzlin, ``The physical meaning of fermi coordinates,'' {\em General
  relativity and gravitation}, vol.~26, no.~6, pp.~619--636, 1994.

\bibitem{chicone2006explicit}
C.~Chicone and B.~Mashhoon, ``Explicit fermi coordinates and tidal dynamics in
  de sitter and g{\"o}del spacetimes,'' {\em Physical Review D}, vol.~74,
  no.~6, p.~064019, 2006.

\bibitem{nesterov1999riemann}
A.~I. Nesterov, ``Riemann normal coordinates, fermi reference system and the
  geodesic deviation equation,'' {\em Classical and Quantum Gravity}, vol.~16,
  no.~2, p.~465, 1999.

\bibitem{muller1999closed}
U.~Muller, C.~Schubert, and A.~E. van~de Ven, ``A closed formula for the
  riemann normal coordinate expansion,'' {\em General Relativity and
  Gravitation}, vol.~31, no.~11, pp.~1759--1768, 1999.

\bibitem{hatzinikitas2000note}
A.~Hatzinikitas, ``A note on riemann normal coordinates,'' {\em arXiv preprint
  hep-th/0001078}, 2000.

\bibitem{yepez2011einstein}
J.~Yepez, ``Einstein's vierbein field theory of curved space,'' {\em arXiv
  preprint arXiv:1106.2037}, 2011.

\bibitem{nuastase2019classical}
H.~N{\u{a}}stase, {\em Classical Field Theory}.
\newblock Cambridge University Press, 2019.

\bibitem{ortin2004gravity}
T.~Ort{\'\i}n, {\em Gravity and strings}.
\newblock Cambridge University Press, 2004.

\bibitem{mironov2014sedeonic}
V.~L. Mironov and S.~V. Mironov, ``Sedeonic equations of
  gravitoelectromagnetism,'' {\em Journal of Modern Physics}, vol.~5, no.~10,
  p.~917, 2014.

\bibitem{kansu2014representation}
M.~E. Kansu, M.~TANI{\c{S}}LI, and S.~DEM{\.I}R, ``Representation of
  electromagnetic and gravitoelectromagnetic poynting theorems in higher
  dimensions,'' {\em Turkish Journal of Physics}, vol.~38, no.~2, pp.~155--164,
  2014.

\bibitem{chanyal2014sedenion}
B.~Chanyal, ``Sedenion unified theory of gravi-electromagnetism,'' {\em Indian
  Journal of Physics}, vol.~88, no.~11, pp.~1197--1205, 2014.

\bibitem{koplinger2007gravity}
J.~K{\"o}plinger, ``Gravity and electromagnetism on conic sedenions,'' {\em
  Applied mathematics and computation}, vol.~188, no.~1, pp.~948--953, 2007.

\bibitem{wald1984general}
R.~M. Wald, ``General relativity(book),'' {\em Chicago, University of Chicago
  Press, 1984, 504 p}, 1984.

\bibitem{schutz2009first}
B.~Schutz, {\em A first course in general relativity}.
\newblock Cambridge university press, 2009.

\bibitem{foster2010short}
J.~A. Foster and J.~D. Nightingale, {\em A short course in General Relativity}.
\newblock Springer Science \& Business Media, 2010.

\bibitem{baldauf2012evidence}
T.~Baldauf, U.~Seljak, V.~Desjacques, and P.~McDonald, ``Evidence for quadratic
  tidal tensor bias from the halo bispectrum,'' {\em Physical Review D},
  vol.~86, no.~8, p.~083540, 2012.

\bibitem{forgan2016tensor}
D.~Forgan, I.~Bonnell, W.~Lucas, and K.~Rice, ``Tensor classification of
  structure in smoothed particle hydrodynamics density fields,'' {\em Monthly
  Notices of the Royal Astronomical Society}, vol.~457, no.~3, pp.~2501--2513,
  2016.

\bibitem{cowles2017cayley}
J.~Cowles and R.~Gamboa, ``The cayley-dickson construction in acl2,'' {\em
  arXiv preprint arXiv:1705.06822}, 2017.

\bibitem{saniga2014cayley}
M.~Saniga, F.~Holweck, and P.~Pracna, ``Cayley-dickson algebras and finite
  geometry,'' {\em arXiv preprint arXiv:1405.6888}, 2014.

\bibitem{manasse1963fermi}
F.~Manasse and C.~W. Misner, ``Fermi normal coordinates and some basic concepts
  in differential geometry,'' {\em Journal of mathematical physics}, vol.~4,
  no.~6, pp.~735--745, 1963.

\bibitem{santos2016introduction}
W.~C.~d. Santos, ``Introduction to einstein-maxwell equations and the rainich
  conditions,'' {\em arXiv preprint arXiv:1606.08527}, 2016.

\end{thebibliography}
\end{document}